# Monitoring lead-acid battery function using *operando* neutron radiography

Jose Miguel Campillo-Robles[1], Damian Goonetilleke[2], Daniel Soler[3], Neeraj Sharma[2], Damian Martin Rodriguez[4], Thomas Bücherl[5], Malgorzata Makowska[6], Pinar Türkilmaz[7], Volkan Karahan[7]

[1] Fisika Aplikatua II Saila, Zientzia eta Teknologia Fakultatea, UPV/EHU, Sarriena auzoa z/g, 48940, Leioa, Basque Country, Spain.

[2] School of Chemistry, UNSW Australia, Sydney NSW 2052, Australia.

[3] Mekanika eta Ekoizpen Industrialeko Saila, Mondragon Unibertsitatea, Loramendi 4, 20500, Arrasate, Basque Country, Spain.

[4] European Spallation Source ESS ERIC, Box 176, 221 00 Lund, Sweden.

[5] Technische Universität München, ZTWB Radiochemie München RCM, Walther-Meißner-Str. 3, D-85747 Garching, Germany.

[6] Paul Scherrer Institut, PSI, 5232 Villigen, Switzerland.

[7] R&D -Engineering Department, Yiğit Akü Malzemeleri A.Ş., Ankara, Turkey.

**Graphical abstract**

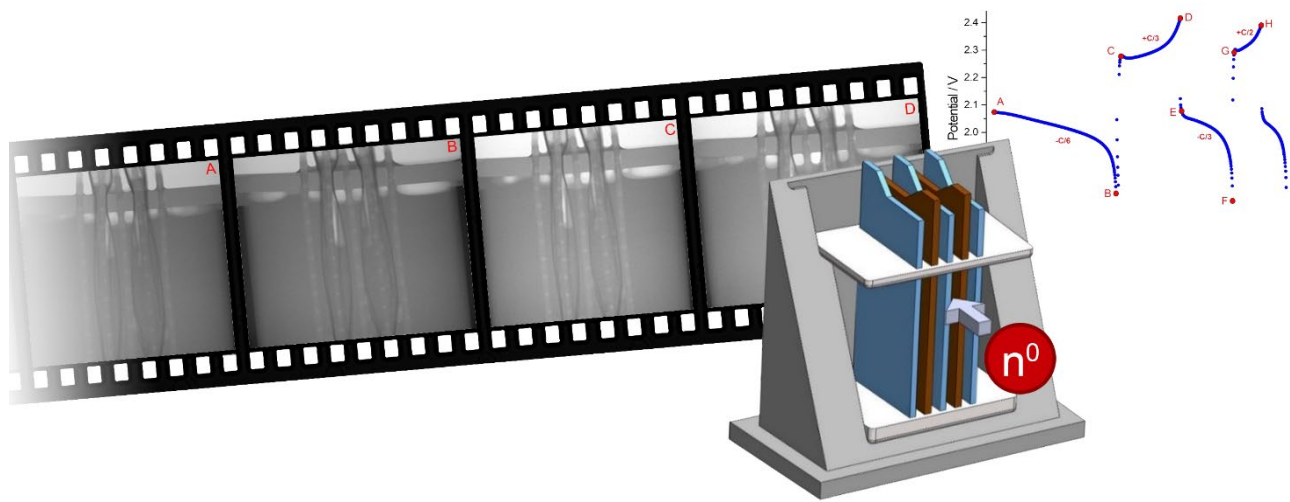

**Abstract**

Investigating batteries while they operate allows researchers to track the inner electrochemical processes involved in working conditions. This study describes the first neutron radiography investigation of a lead-acid battery. A custom-designed neutron friendly lead-acid cell and casing is developed and studied *operando* during electrochemical cycling, in order to observe the activity within the electrolyte and at the electrodes. This experimental work is coupled with Monte Carlo simulations of neutron transmittance. Details of cell construction, data collection and data analysis are presented. This work highlights the potential of neutron imaging for tracking battery function and outlines opportunities for further development.

**Research Highlights**

- First neutron radiography of *operando* lead-acid batteries is reported.
- Neutron friendly lead-acid battery is designed and manufactured for experiments.
- Neutron transmittance of electrodes increases/decreases during discharge/charge.
- Neutron transmittance measurements confirm Monte Carlo simulations.

# 1. Introduction

Rechargeable or secondary batteries have been established as essential components in a variety of important applications, ranging from lithium-ion batteries in personal electronic devices and traction of electric vehicles to peak load levelling devices in smart grids. Lead-acid batteries, unarguably the original rechargeable battery technology developed by French physicist Gaston Plante in 1860 [1], are still finding specialised use due to their low cost and proven reliability [2, 3]. These applications include starters and portable power for vehicles, submarine power, storage of excess energy from renewable sources and backup or uninterrupted power supply (UPS) systems, among others [4, 5]. The basic principles of lead-acid battery operation have not diverged from the originally reported mechanism, involving the diffusion of different ions through the sulphuric acid-water electrolyte between a porous lead negative electrode and a porous lead oxide ($PbO_2$) positive electrode [6]. However, the efficient recycling of toxic lead and innovations in the cell construction have kept the technology pertinent and enhanced its usefulness for new applications. From the point of view of construction, various types of lead-acid batteries exist, including flooded batteries, sealed batteries or commonly named valve-regulated lead-acid batteries (VRLA), absorbed glass mat sealed batteries (AGM) and gel sealed batteries. Lead-acid batteries can function as deep cycle batteries, cranking batteries, standby batteries and marine batteries, to name a few. More recently, researchers have reported interesting developments in lead-acid technology. For example, zinc-lead dioxide cells which can operate at higher potentials have been developed [7]. Improved electrode and electrolyte additives have been used to obtain higher performance lead-acid cells [8]. A lead-acid metal hydride hybrid redox couple battery has been developed using three electrolytes [9]. Finally, a lead-acid battery and a supercapacitor have been hybridised into one device, in examples such as the commercial Ultrabattery$^{TM}$ [10].

Considering such a wide variety of applications and high stability of this technology, it would be expected that available characterisation is fairly comprehensive. However, there are few reports of techniques for monitoring the internal processes of the lead-acid batteries. The nature of the lead-acid battery, full of large lead electrodes, corrosive sulphuric acid and variable electric currents that occur during operation, makes it difficult to perform an inner *operando* monitoring. While most studies have focused on *ex situ* or *post mortem* characterisation, a number of *in situ* studies have been undertaken [11-14]. Usually the main objective is the measurement of the density/concentration of the electrolyte to analyse, for example, the state of charge (SOC) of the cell or the stratification process. Different techniques have been used for this purpose (see for example [15] for a review). Most researchers have used optical techniques [16-19]. The physical principle involved is the linear relationship between the specific gravity and the refractive index of the electrolyte. Another commonly used technique is the measurement of the equilibrium potential [20, 21]. Unfortunately, these techniques are intrusive; i.e. they need to insert sensors in the cell, thus changing its behaviour. To our knowledge, four non-intrusive techniques have been used to monitor lead-acid batteries. First, some researchers have measured the electrolyte specific gravity/density using ultrasound [22, 23]. Second, holographic laser interferometry (HLI) has been used for continuously monitoring of lead-acid batteries during charge/discharge cycling [24-26]. Another interesting and recent technique is the use of magnetic field measurements to examine current distributions within the cell [11]. Finally, synchrotron X-ray radiography and tomography has been used to follow 3D changes in morphology with μm resolution [27].

The use of neutron and X-ray radiography to study energy storage devices has been successfully demonstrated for various types of lithium-ion battery chemistries [28-32] and fuel cells [33-39]. For example, neutron radiography has been used to show macroscopic information such as electrolyte distribution inside a lithium-ion battery [28] or lithium distribution within a LiFePO$_4$ pouch cell [29]. Positive electrode material (PbO$_2$) of lead-acid batteries has been analysed previously using neutron transmission, inelastic neutron scattering [40] and neutron powder diffraction [41]. Lead-acid batteries have been analysed using X-ray radiography (*post mortem*) [42] and synchrotron radiography and tomography (*operando*) [27]. However, neutron imaging has not been used previously to study lead-acid batteries *operando*. The high penetration depth of neutron radiation through most materials [35] makes it ideal for visualising the processes occurring within large devices, if a suitable vessel to contain the device components without impeding the neutron beam is designed. As the strength of the scattering cross section is non-uniform across the periodic table for thermal neutrons, and depends on particular isotopes of each element [43], careful selection of materials when designing the cell components is essential to ensuring a good signal-to-noise ratio.

This manuscript explores the viability of neutron imaging to obtain measurable information of the internal processes in lead-acid batteries. A lead-acid battery that is suitable for *operando* neutron imaging studies has been designed and demonstrated. This paper illustrates the design of this cell optimized for neutron studies and shows the first results of experiments conducted at thermal neutron imaging facilities: DINGO (ANSTO, Sydney) [44, 45] and NECTAR (FRM II, Garching) [46-48]. This research, to the best of our knowledge, presents the first reported study of applying neutron imaging techniques to understand the working processes of lead-acid batteries.

## 2. Experimental

### 2.1. Cell Design

Although conventional lead acid batteries use Acrylonitrile butadiene styrene (ABS) or Polypropylene (PP), custom cell casings were designed and manufactured from polystyrene and Teflon [49]. These materials were chosen as they offer good chemical resistance to the concentrated sulphuric acid used in the electrolyte, they can be easily manufactured, and most importantly minimise hydrogen content to maximise the neutron signal from the materials inside the casing. The Teflon casing was found to offer better neutron transmittance, which can be attributed to fact that the material does not contain any hydrogen. The schematic of the cell casings are depicted in Figures 1 a) and b). The maximum length of the polystyrene casing in the direction of the neutron beam is 30 mm ($L$), while its perpendicular section to the beam is 75 mm ($H$, height) × 81 mm ($W$, width) (see Figure 1 a)). The Teflon casing has dimensions of 31.4 mm ($L$) × 75 mm ($H$) × 60 mm ($W$) (see Figure 1 b)). The dimensions were chosen such that they provide sufficient path length for neutrons to traverse from the source to the detector and such that the detector field of view was slightly bigger than the case (neutron field of view greater than $H \times W$). The walls of the case perpendicular to the neutron beam were ground to a thickness of 0.5 mm in the polystyrene case and 0.7 mm in the Teflon case, while the edges parallel to the neutron beam remain thicker to provide rigidity to the structure. The distance traversed by the neutron beam was 30 mm for the polystyrene cell and 31.4 mm in the Teflon cell.

Lead-acid batteries typically use porous lead as the negative electrode and porous lead oxide, ($PbO_2$) as the positive electrode. These active materials are very brittle. As a result, they need a current collector of metallic lead, which gives the structural rigidity to the electrode. In this study, standard dry charged electrodes with a mean thickness of 1.7 mm were supplied by YIGITAKU (see Supplementary Information). The positive and negative electrodes were manually cut down, using scissors, to a size that fits into the custom designed battery, approximately 26 mm × 60 mm (without tab) for polystyrene case. During the cutting process, some electrodes were damaged because active mass is very brittle, and those electrodes were removed. Indeed, it is a very difficult process and many electrodes were discarded. This configuration provides an overall capacity of 1.2 Ah (at C/20). In order to ensure the electrodes are aligned parallel to the neutron beam, two horizontal electrode guides (made of Teflon) are inserted into the case, at the top and bottom of the case (see Fig. 1 a) and b)). Guide pairs were manufactured to achieve equal spacing between the electrodes. A five electrode configuration with 3 negative electrodes and 2 positive electrodes was used (see Fig. 1 c) and Supplementary Information). Electrical connections between electrodes were built using screws and nuts (see Fig. 1 d) and Supplementary Information). A deuterated electrolyte solution was used, consisting of 5 M of deuterated sulphuric acid $D_2SO_4$ in deuterated water $D_2O$ (see Table 1). This is necessary as hydrogen has a large incoherent neutron-scattering cross section, which would result in poor signal-to-noise ratio in the data [50]. Deuterated electrolyte has been used previously to perform neutron scattering measurements on lithium-ion batteries [51-53]. However, as the heavier deuterium atom is active in the redox reaction in the lead-acid system, it could possibly result in slower deuterium diffusion and slower redox reaction kinetics [54]. No significant adverse consequences of using deuterated electrolyte were observed in this study, however directly comparable cells using conventional electrolyte were not investigated. The electrolyte was prepared in an Ar-filled glovebox and stored in a sealed sample container. Electrolyte was added to cells in a glovebox 2-3 minutes prior to transport to the neutron imaging instrument.

### 2.2. Neutron imaging instrumentation

Neutron imaging was carried out using two instruments to probe subtly different aspects and applicability on multiple instruments. One set of experiments was undertaken using DINGO, the neutron imaging instrument situated at the OPAL reactor at the Australian Nuclear Science and Technology Organisation (ANSTO, Lucas Heights, Australia) [44, 45]. The instrument offers a neutron flux of up to $4.75 \times 10^7$ $cm^{-2}$ $s^{-1}$ at the sample position, with a wavelength distribution centred at 1.08 Å. Table 2 shows the principal technical characteristics of DINGO. In these experiments, images were captured on an area of 10 cm × 10 cm, and they were collected every minute with a 45 s exposure time. The $L/D$ relation was 1000, and the CCD sensor produced a pixel size of 26.8 μm. Another set of experiments was conducted using NECTAR, the NEutron Computed Tomography And Radiography instrument, which is located on the FRM II reactor operated by the Technical University of Munich (TUM, Garching, Germany). NECTAR is the only facility that offers fission and thermal neutrons together for 2D and 3D imaging [46-48]. The principal technical characteristics of NECTAR appear in Table 2. In our measurements, images were captured on an area of 30 cm × 30 cm, and they were collected every 20 s with a 15 s exposure time with an *L/D* relation of 230. The detector setup was composed of a 200 μm thick $^6$LiF / ZnS:Cu scintillator and an Andor iKon-L CCD camera with pixel size of 13.5 μm.

### 2.3. Neutron imaging electrochemistry

Several lead-acid batteries of 1.2 Ah capacity (nominal voltage 2.1 V) were built using the polystyrene casings and deuterated electrolyte was added to the cells 2-3 minutes prior to transport to the neutron instruments. Prior to cycling, the dry charged electrodes were subjected to an activation cycle (during which neutron images were also collected). The activation protocol was 20 min resting period such that the electrodes are at about 85 % state of charge (SOC), galvanostatic (constant current) 2 A discharge until 1 V and potentiostatic (constant voltage) 2.67 V (±0.017) during 24 h, with a maximum current of 0.3 A. After the activation, the electrochemical cells were cycled in galvanostatic mode between C/6 (0.156 A) to C/2 (0.380 A) using an Autolab potentiostat/galvanostat (PGSTAT302).

### 2.4. Simulations

Monte-Carlo simulations were performed to understand the experimental trends of neutron transmittance. Particle and Heavy Ion Transport code System, PHITS, was used to model the transmittance of the cell and electrolyte, and also to model the sensitivity of neutron imaging to changes in electrolyte concentration [55, 56]. The PHITS code uses the Monte-Carlo method to simulate radiation transport processes. It can simulate transport processes of most of the particle species (neutron, electron, protons, photons, etc.) with energies up to 1 TeV (per nucleon for ion) by using several nuclear reaction models and data libraries. The simulated system is illustrated in Fig. 2 a). A disk shaped source emits neutrons with an incident energy of 28 and 100 meV, respectively. The cell is placed 6 m away from the source and the beam profile is measured at 6.5 m from the source. The geometry of the cell used in the simulation is a simplified version of the one described in Fig. 1. The cell with its corresponding materials appears in Fig. 2 b). In the case of evaluating the transmittance of the electrolyte, the electrodes have been removed, leaving only $D_2O$-$D_2SO_4$ electrolyte inside the simulated cell.

## 3. Results and Discussion

### 3.1. Case

After manufacturing, the empty polystyrene case was tested with the thermal neutron beam, and the measured mean transmittance was 75.3 ± 4.6 % at DINGO (see Fig. 3) and 71.3 ± 2.6 % at NECTAR. Comparatively the Teflon case produced a mean neutron transmittance of 96.4 ± 1.7 % at NECTAR. Fig. 3 shows the corrected image of the empty polystyrene case (see correction procedure in Supplementary Information). Manufacturing defects that reduced the thickness of the 0.5 mm walls can be observed as two vertical orange lines in both extremes of the case (see Fig. 3). Moreover, a fold of one or both walls of 0.5 mm appears as the inner orange region. This highlights the difficulties with manufacturing with polystyrene and the small wall thicknesses.

### 3.2. Electrolyte

Radiography data was collected from a cell casing containing no electrodes and filled with varying concentrations of electrolyte. This was performed to check if it is possible to measure electrolyte concentration changes inside the battery using neutron imaging. These concentration dependent radiography scans were further compared with simulations of neutron transmission through the custom battery.

In working conditions, the mass fraction of sulphuric acid in the electrolyte of lead-acid batteries usually ranges between 10 and 40 wt. % [2]. In these experiments, a deuterated version of the electrolyte was used, consisting of deuterated sulphuric acid, $D_2SO_4$, and heavy water, $D_2O$, as the cross-section for neutrons of hydrogen is much higher than of deuterium, thus increasing the sensitivity drastically. Although, many properties of $D_2O$ have been analysed [57-60] and some of the properties of $D_2SO_4$ too [61, 62], the properties of $D_2SO_4$-$D_2O$ electrolyte are not well known yet [63, 64].

The neutron transmittance of the deuterated electrolyte was measured at various concentrations using both the DINGO and NECTAR neutron imaging instruments. Before the neutron measurements, all samples were stirred for at least 5 minutes to obtain a uniformly mixed electrolyte. The correction procedure explained in the supporting information was applied to the measured data removing the effects of the case. The neutron transmittance was averaged over the electrolyte region to evaluate the attenuation coefficient. In this calculation, the limiting regions of the electrolyte were the thick walls of the case which were removed to avoid the stepwise change of the neutron intensity (see white region in Fig. 4 a)).

At DINGO, pure deuterated sulphuric acid was dropped into a polystyrene case (see section 2.1 and Fig. 1 a)). Afterwards, pure deuterated water was slowly added to reduce the sulphuric acid concentration from 100 wt. % to 20.00 wt. %. The transmittance of pure heavy water, as the limiting point in the concentration range (0 wt. %) was also measured.

A similar procedure was followed at NECTAR, but using a Teflon case (see section 2.1 and Fig. 1 b)). The analysed concentrations ranged from 100 wt. % to 25.03 wt. %, and, in order to obtain better quality data in the low concentration region, the measurements were repeated starting from pure heavy water and adding deuterated sulphuric acid up to a concentration of 48.15 wt. %.

All measured transmittances appear in Fig. 4 b). Error bars were computed using standard deviation of transmittance of measured area with a cover factor of 2 [65]. The neutron transmittances obtained at DINGO are larger than those measured at NECTAR (see Fig. 4 b)). This can be explained partially because of the small difference in thickness of the two samples (DINGO: 29 mm,

NECTAR: 30 mm). Moreover, the resultant neutron attenuation coefficients depend on the neutron spectrum, and higher energy spectrum results in larger transmittance of neutrons [35]. Fig. 4 b) shows that neutron transmittance remained constant in the low concentration region, which is very important and slightly unfortunate, because in a lead-acid battery the expected electrolyte concentrations fluctuations are in this region (10-40 wt. %). Therefore, in the working battery, it is difficult to measure concentration change of the electrolyte using neutron imaging (unless the electrolyte becomes over-concentrated). In the high concentration range, the neutron transmittance increases with $D_2SO_4$ concentration towards the value of pure $D_2SO_4$. Furthermore, in measurements at NECTAR, there are small discrepancies between the two experimental procedures (dilution approach) in the 25.03-48.15 wt. % range (see Fig. 4 b)). This can be explained because $D_2O$ is more likely to evaporate than $D_2SO_4$ in the very exothermic mixing procedure (vapour pressure of water is three orders of magnitude greater than sulphuric acid). For this reason, the high concentration curves do not perfectly match each other. Nevertheless, these minor discrepancies are within the error range of the measurements. This is a classic chemical example of an open system.

From these measurements, we have also obtained the linear attenuation coefficient of the electrolyte using Equation (S2) of the supporting information and the error bars (see Fig. 4 c)). The coefficient decreases as the concentration of the electrolyte increases, and this trend is similar in both measurements with both cases and instruments. The coefficients obtained at NECTAR are higher than those of at DINGO, most likely related to the different spectrum of the incident beam. Moreover, different cells have been used in each facility and they scatter neutrons differently, so they can change the absolute measured values as well. This effect should be very small after normalization with an empty cell, but a completely direct comparison of these experiments should take these slight differences into account. In Table 3 and Fig. 4 d), we have collected our measured results for $D_2O$ and $D_2SO_4$ and other values obtained from the literature [66, 67]. All the values show the same linear trend, in which the neutron attenuation of the deuterated water is larger than that of the deuterated sulphuric acid.

The PHITS code was implemented with the experimental geometrical arrangement and Fig. 5 shows the most important results of these simulations. The neutron transmittances of $D_2O$ and $D_2SO_4$ were calculated for different thicknesses (see Fig. 5 a)). The neutron attenuation coefficient of $D_2O$ is larger than for $D_2SO_4$, and the deuterated electrolyte transmittance should have a value between those of $D_2O$ and $D_2SO_4$. These results are in accordance with the measurements performed at DINGO and NECTAR. In addition, Fig. 5 b) shows, the simulation of the cell with different electrolyte concentrations assuming two different values of energy of the incident neutron beam: 28 and 100 meV. The transmittance shown in Fig. 5 b) has a monotonous increment with the concentration, and as expected, the beam of 100 meV has a greater transmittance than that of 28 meV.

### 3.3. Electrodes

The transmittance of the electrodes was also investigated. Commercial electrodes were cut and the neutron transmittance of these dry charged electrodes was measured at DINGO without any electrolyte. In the upper part of the case, the transmittance is reduced due to the horizontal guide of Teflon added to maintain the electrodes parallel (this guide is not in the empty case image, Fig. 6 a)). Distinct yellow points of greater intensity can be seen inside all the electrodes. These points represent the current collector grid made of metallic lead that is used to hold the active material in place. In Fig. 6 b), two guidelines are represented vertical (blue) and horizontal (black). The neutron transmittance change through these lines has been plotted to see the differences between materials. The vertical blue line shows the neutron transmittance along the second electrode from the left, a positive electrode (see Fig. 6 c)). Metallic lead points show greater transmittance than the active material, porous lead oxide ($PbO_2$). The transmittance of these grid points is greater than 50 %, and that of the porous lead oxide active material is lower, showing a curved profile. A similar profile is obtained for the other electrodes, independently of whether they are positive or negative electrodes. The horizontal black line shows the neutron transmittance of all the electrodes (see Fig. 6 d)). The transmittance in this line shows two abnormal peaks in the two sides of the case (in opposite directions) that is originated by a small movement of the case during the insertion of the electrodes. All the electrodes show a similar transmittance around 50 %. However, the first electrode of the right side has a broadened peak due to a slight non-parallel alignment (see Fig. 6 b) and d)).

Simulations of the effect of porosity on the neutron transmittance of electrodes were performed using PHITS code, but the effect of the electrode grid was not taken into account. Porosity has been modelled as a reduction of the average density of the electrode, from 0.0 to 0.8 porosity. If there is no porosity (0.0 case), both electrodes show the same neutron transmittance, and as expected simulated transmittance increases with increasing porosity. However, the neutron transmittance of the negative electrodes (with higher density and lower oxygen content) has a lower transmittance than the positive electrodes, which could indicate the influence of oxygen content on the neutron transmittance.

### 3.4. *Operando* neutron radiography

Initially, cells were constructed using separators around the positive electrodes, i. e. the second and fourth electrodes (see Fig. 7 a) and b)). The cell is designed to have 2 mm between all the surfaces of separators and negative electrodes. *Operando* neutron imaging was carried out initially at DINGO and the correction procedure described in the supporting information was applied to the measured data. The imaging data can clearly resolve the electrodes, separators, electrolyte and case components. Fig. 7 a) shows that separators were not parallel to electrodes and were bent, because it was very difficult to hold the separator to the surface of the positive electrodes. Moreover, we observed the evolution of gas in the electrolyte/electrode surface (see red regions between electrodes in Fig. 7 b)). Gas evolution has been measured in operating lithium-ion batteries using neutron imaging [68, 69]. In lead-acid batteries, this process appears predominantly in the potentiostatic charge of the activation process where the battery is held at 2.67 V (greater than the gassing voltage, 2.4 V) [2]. This correlates to the fact that when a lead-acid battery is overcharged, $O_2$ is evolved at the positive plate and $H_2/D_2$ is evolved at the negative plate [70-72]. The gas generated in the activation travels along the electrode/separator surface upwards. The separators appear to seed the generation of bubbles, a surface on which the bubbles can grow and be transported from the bottom of the cell (or at any level) to the liquid surface level. As mentioned, the separators were not parallel to the neutron beam throughout the vertical profile, so they were removed in subsequent iterations of the cell.

The removal of separators from the cell allows the electrodes and regions of electrolyte to be better resolved (see Fig. 2 d) and 7 c)) and did not have any adverse effects on the electrochemical behaviour of the cell. However, in order to avoid short circuits, the distance between the surfaces of all electrodes was increased to 4 mm. Some galvanostatic charge-discharge processes have been applied to the lead-acid battery to investigate the effects that electrochemical processes have in the neutron transmission (see Fig. 8 a) and Table 4). During lead-acid battery discharging process, $PbSO_4$ precipitates in/on both electrodes [2]. As a result, the porosity of electrodes is reduced. Moreover, the electrolyte concentration reduces in the cell, but principally near the electrodes. However, in the positive electrodes, the concentration reduction of the electrolyte is greater than in negative electrodes (see Supplementary Information and Ref. 73). The opposite happens during charging processes. Neutron transmittance of a horizontal line of the cell was obtained during all the electrical processes of Fig. 8 a). The initial process is a galvanostatic discharge (see Table 4 and Fig. 8 a)), with a discharging current of C/6 (0.156 A) and a discharged capacity of 1.18 Ah for 7.6 h. During this discharge, the neutron transmittance of the three inner electrodes shows a slight increment (see Fig. 8 b)). Fig. 8 c) shows the mean neutron transmittance over a small area of the electrodes (three negative and two positive between them). During the second process, a galvanostatic charging of C/3 (0.272 A), neutron transmittance of all the electrodes has a slight reduction (see Fig. 8 b) and c)). Moreover, in the other galvanostatic processes of Fig. 8 a), neutron transmittance shows the same behaviour (see Fig. 8 b) and c)). These changes of neutron transmittance cannot be explained with the change of concentration generated in the electrolyte during charging-discharging. This variation may be attributed to the change of chemical composition of the positive electrodes (from Pb/$PbO_2$ to $PbSO_4$, and backwards). $PbSO_4$ has higher molar volume than the other electrode materials. Therefore, during discharging process the porosity of the electrodes reduces (the liquid fraction of the electrodes reduces) [74]. Finally, in Fig. 8 b), neutron transmittance of the two outer negative electrodes shows the same behaviour, but its value is smaller than that of the three inner electrodes. These outer electrodes may have less electric current due to the five-electrode configuration, and further have two edges that are not facing an opposite electrode. As a result, the electrochemical reactions that take place in/on them happen slowly, and the amount of $PbSO_4$ (Pb and $PbO_2$) created in them during discharging (charging) is lower.

## 4. Conclusions

The first neutron transmission measurements of a lead-acid battery have been demonstrated. Neutron friendly polystyrene and Teflon cases have been designed and manufactured to perform the measurements. The optimisation indicates that the Teflon case can be built more reliably and shows better data quality (a case of PP monomer might be an alternative in future experiments further reducing the hydrogen content with respect to PS). Experiments show that it is difficult to observe any change of electrolyte concentration during conventional operation using neutron imaging, because there is not sufficient contrast. However, transmission measurements indicate that electrochemical changes in the electrodes can be detected. More experiments are necessary to resolve better the differences observed in neutron transmission during electrochemical processes and this work lays the foundation for future investigations.

## 5. Acknowledgements

We acknowledge the support of the ANSTO, as well as, Dr Ulf Garbe and staff at ANSTO for assistance during the experiment. Authors are also grateful to I. Urrutibeaskoa (MU), F. Zugasti (MU) and E. Ruiz de Samaniego (Fagor Automation) for their technical help in the design of the battery cases and Parra Mekanizatuak for manufacturing the cases. Authors wish to recognize students James Christian (UNSW), Junnan Liu (UNSW) and Jimmy Wu (UNSW) for their assistance running the DINGO experiments.

# Figure captions

**Figure 1.** a) Dimensions of the polystyrene case (in mm), b) dimensions of the Teflon case (in mm), c) inner structure of the neutron friendly battery cell without separators (colour code: negative electrodes, blue; positive electrodes, brown). The arrow indicates the direction of the incident neutron beam. d) neutron friendly battery with polystyrene case ready to be cycled.

**Figure 2.** a) Sketch of the geometrical arrangement of the simulations performed with PHITS code ($L/D = 500$). b) Sketch of the battery cell used in the simulations performed with PHITS.

**Figure 3.** Neutron transmittance of one of the polystyrene cases (DINGO). Greater transmission zones in the inner part of the case show some manufacturing faults.

**Figure 4.** a) NECTAR measurement of the transmittance of the electrolyte. The mean value of the transmittance has been calculated in the white area. b) Measured neutron transmittance of the sulphuric acid-water electrolyte in all the concentration range (DINGO: triangles, NECTAR: circles). c) Measured linear attenuation coefficient of the electrolyte as a function of the concentration (DINGO: triangles, NECTAR: circles). d) Linear attenuation coefficients of deuterated water (1.107 g cm$^{-3}$) and sulphuric acid (1.86 g cm$^{-3}$): measured at DINGO, measured at NECTAR and NIST [67] (DINGO: triangles, NECTAR: circles, NIST: squares).

**Figure 5.** Neutron transmittance simulations performed with PHITS code: a) different thicknesses of D$_2$O and D$_2$SO$_4$ inside the Teflon case. b) All the concentration range of the electrolyte (inside the case) for incident neutrons of 28 and 100 meV energy.

**Figure 6.** Neutron transmittance images of: a) empty cell used to analyse the electrodes and b) electrodes placed at the polystyrene case (DINGO). Neutron transmittance plot of: c) blue vertical line and d) black horizontal line. e) Simulated neutron transmittance of the electrodes inside the battery case as a function of the porosity of the electrodes.

**Figure 7.** Neutron transmittance images of 1.2 Ah lead-acid battery cell (2 mm separation): with separators a) as measured and b) after data reduction, and without separators c) as measured and d) after data reduction.

**Figure 8.** a) Measured battery voltage of a galvanostatic C/6 discharge, 0.156 A (4 mm separation). b) Neutron transmittance evolution of a horizontal line that crosses the entire cell (height of measurement: 1660 pixel) during all the electrical processes that happens in Fig. 8 a). c) Mean neutron transmittance evolution of a small area of the electrodes (height of measurement: 1660 pixel) during all the electrical processes that happens in Fig. 8 a).

**Figure 1**

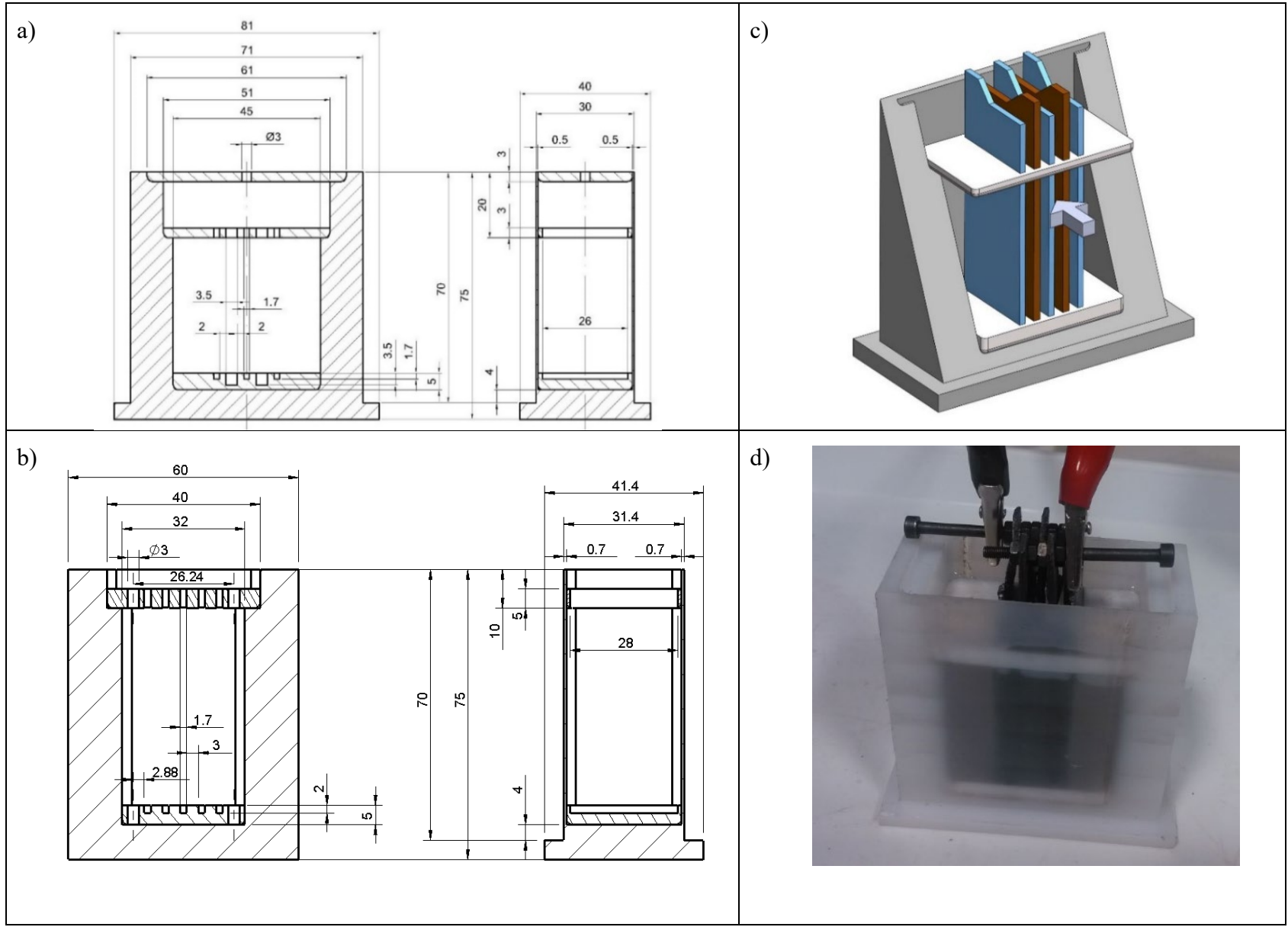

**Figure 2**

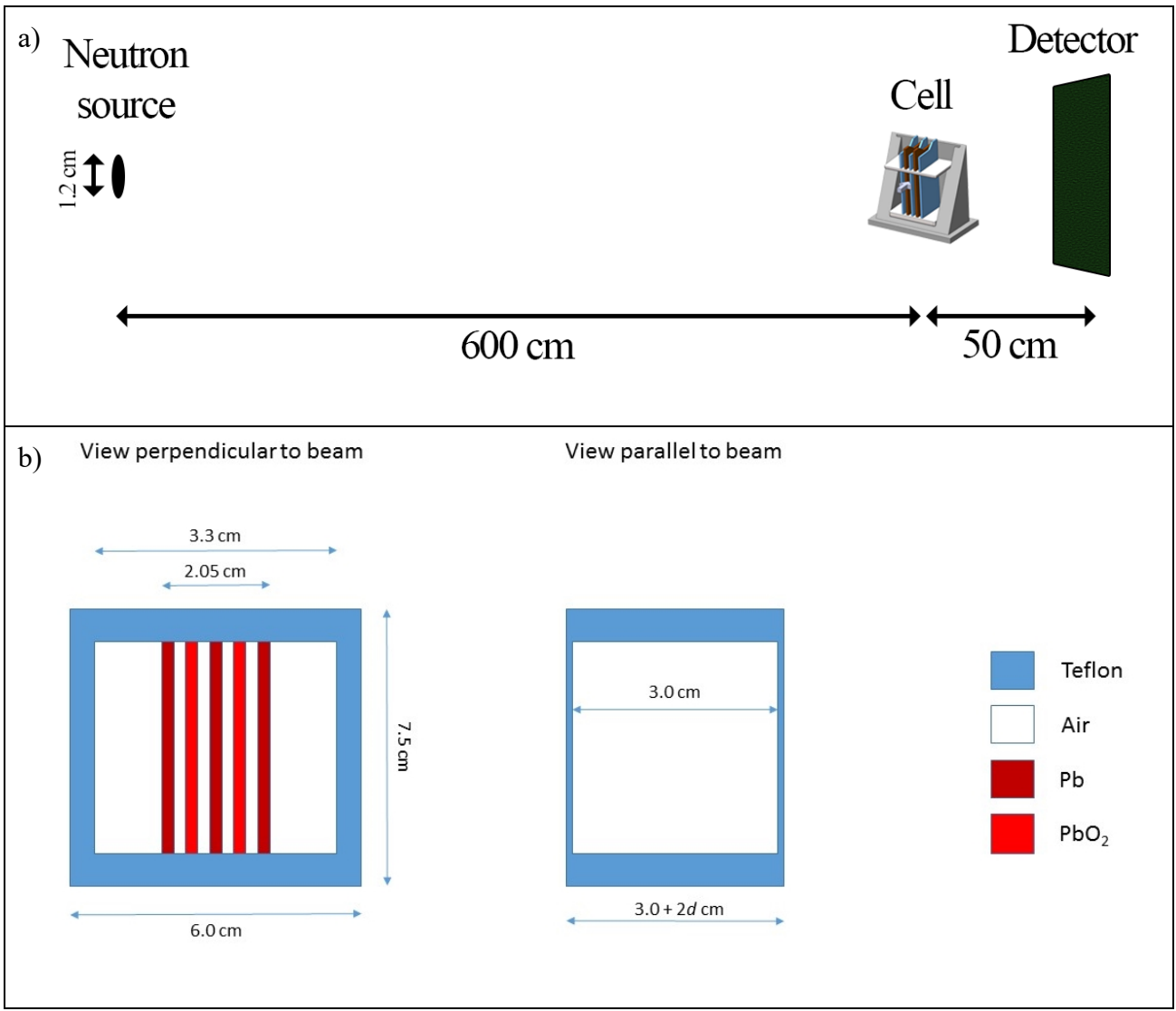

**Figure 3**

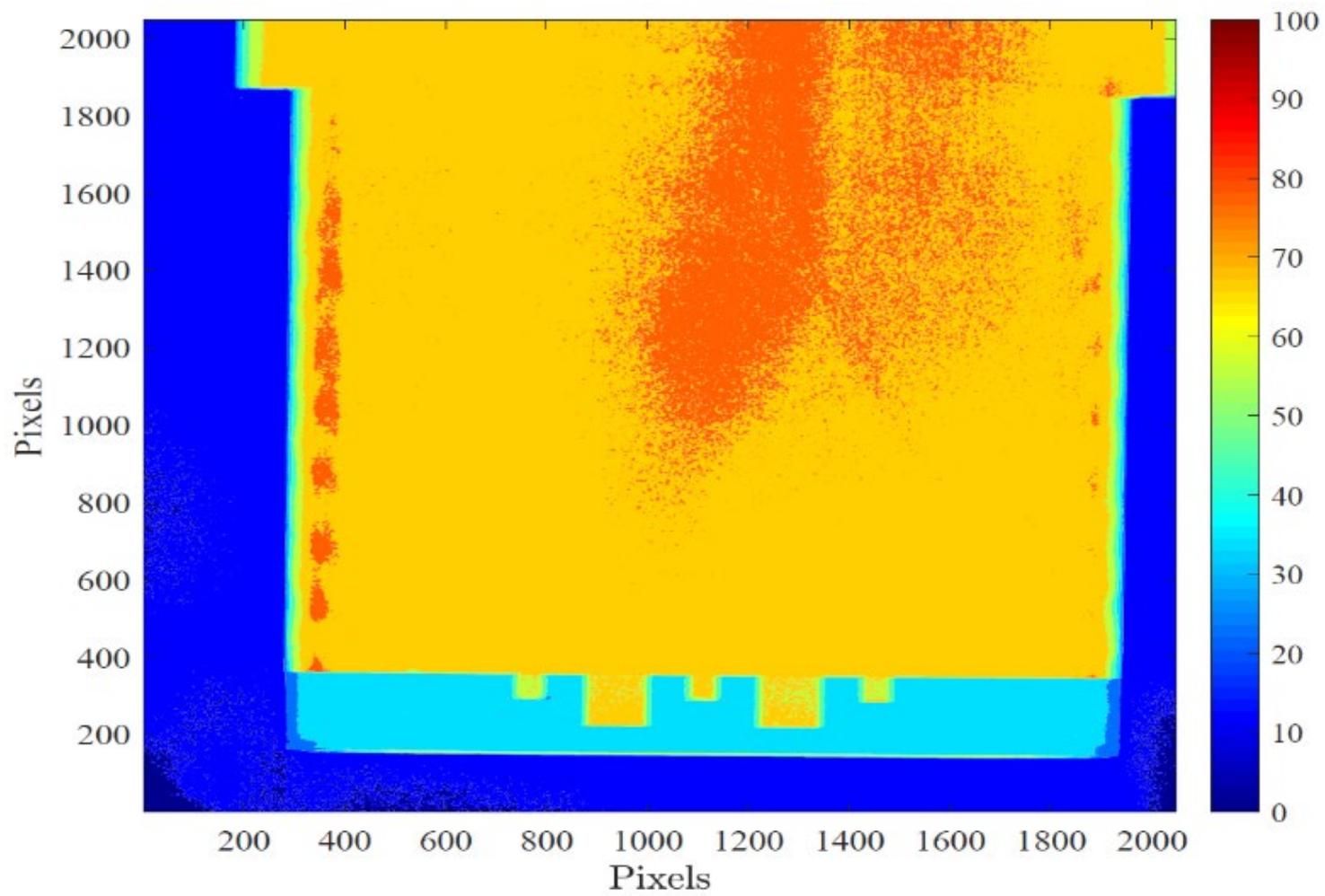

**Figure 4**

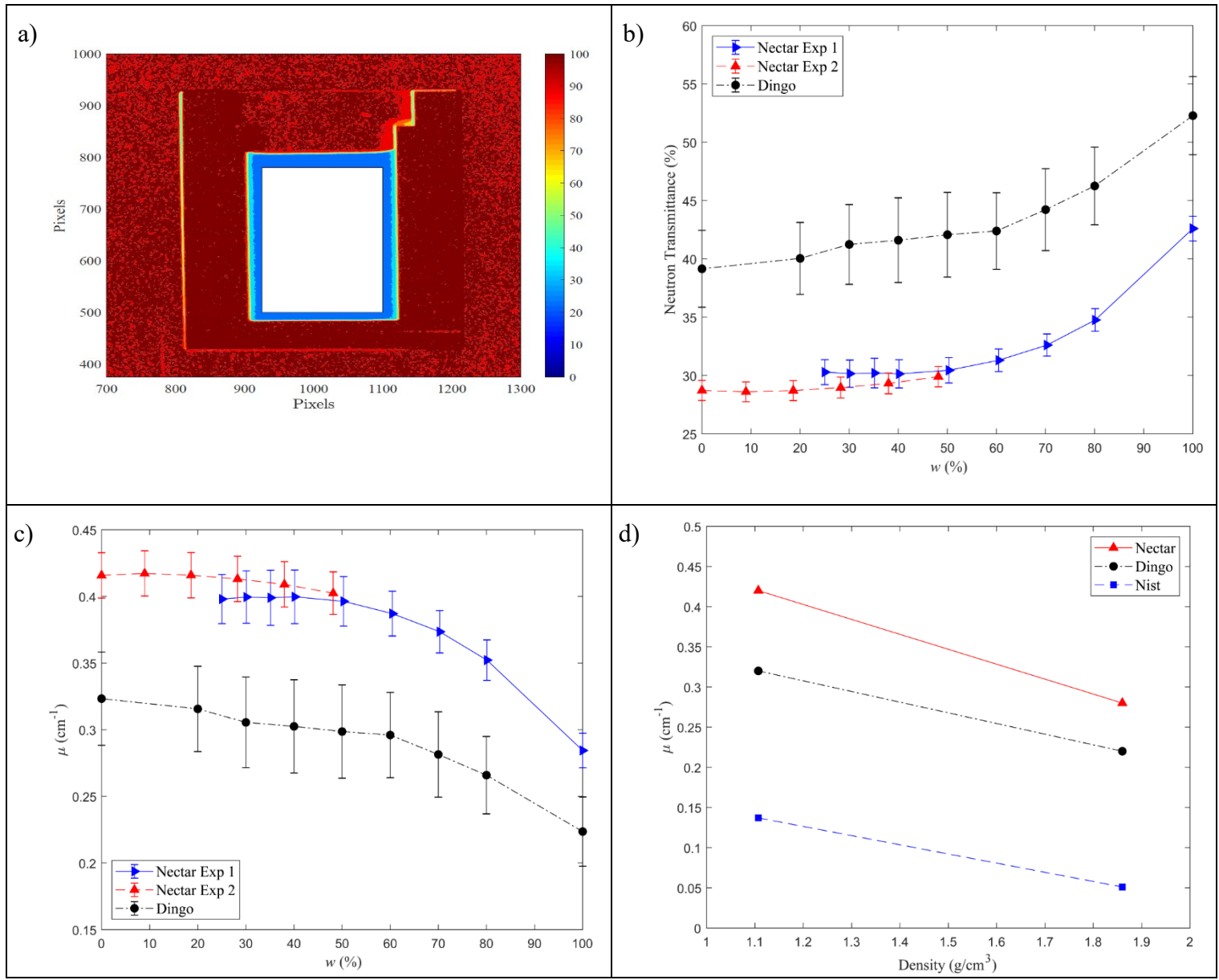

**Figure 5**

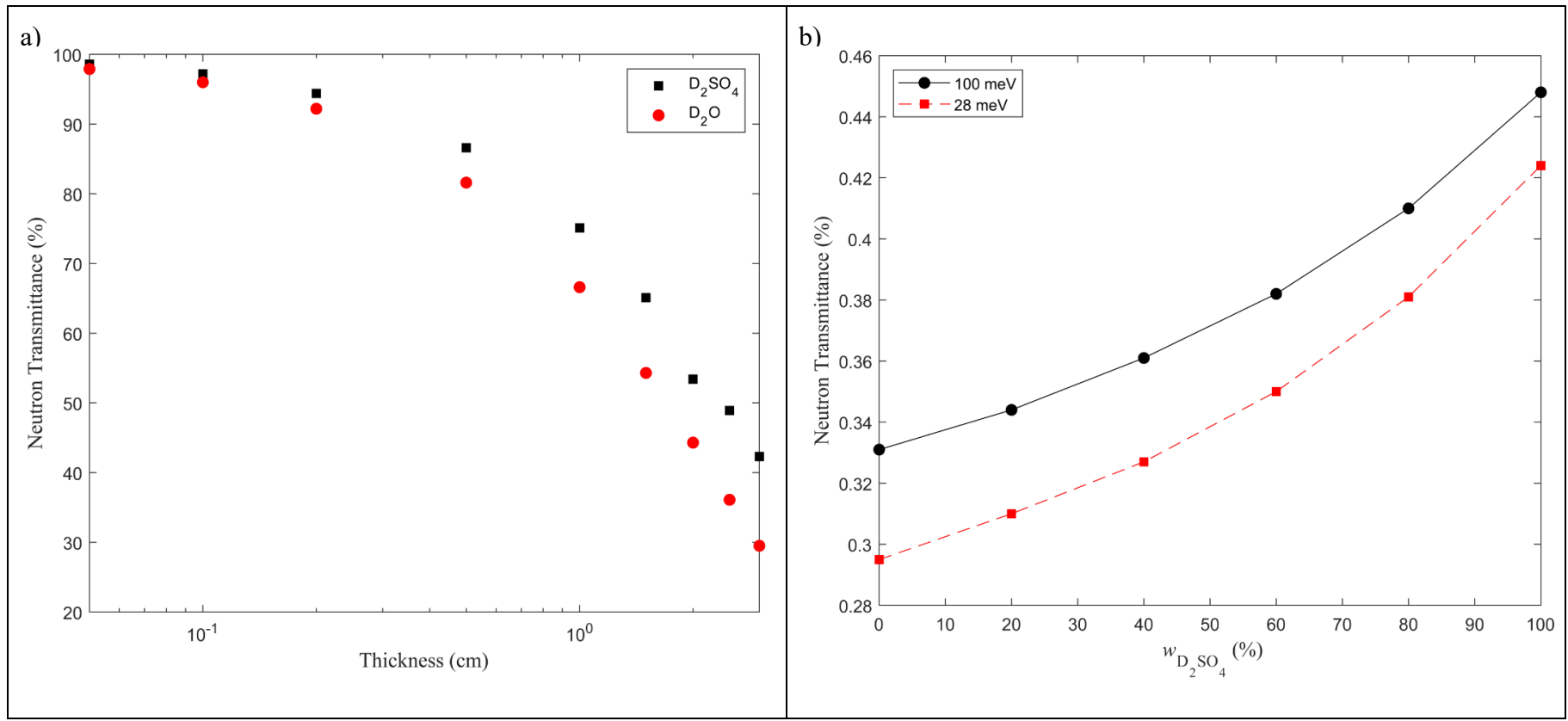

**Figure 6**

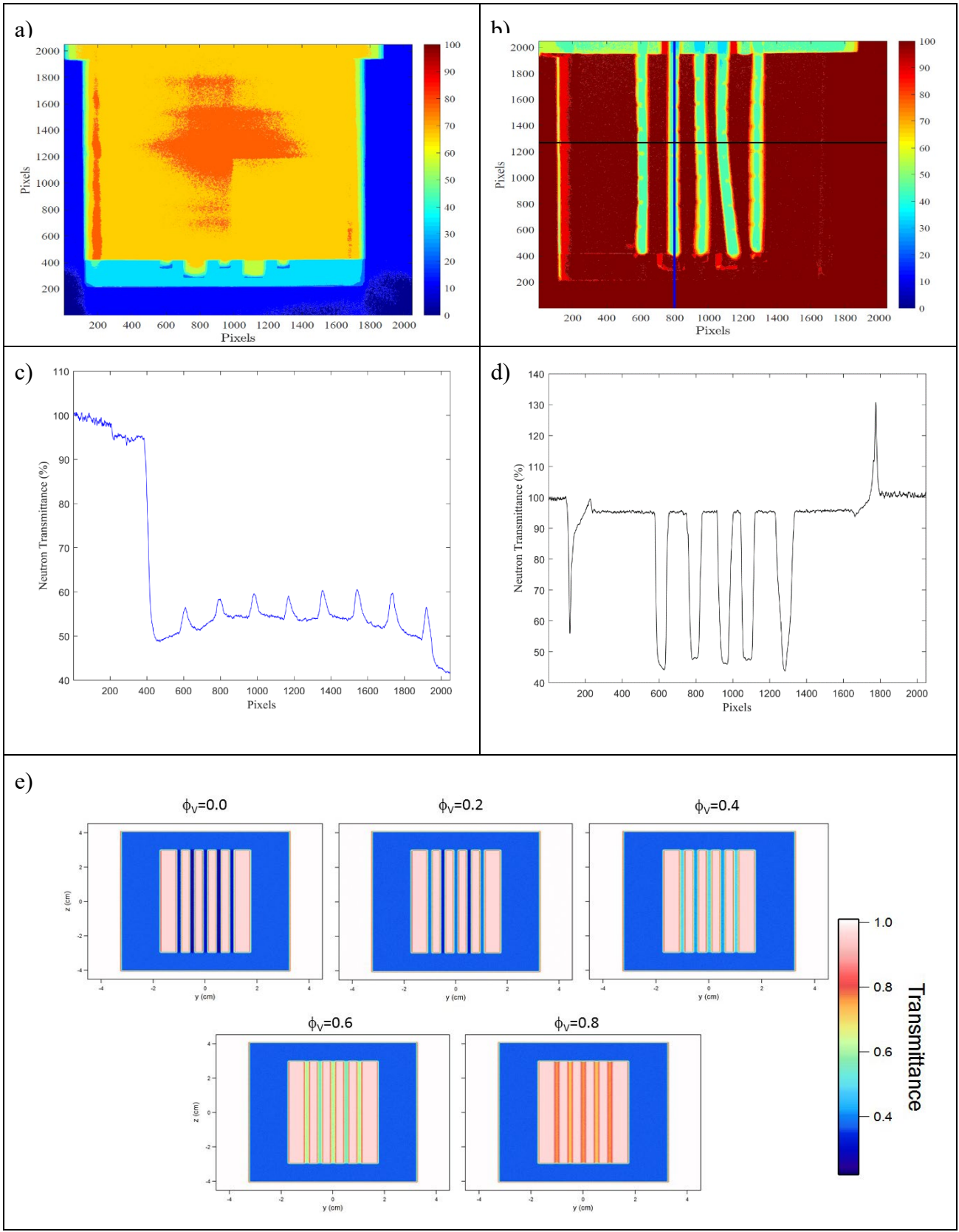

**Figure 7**

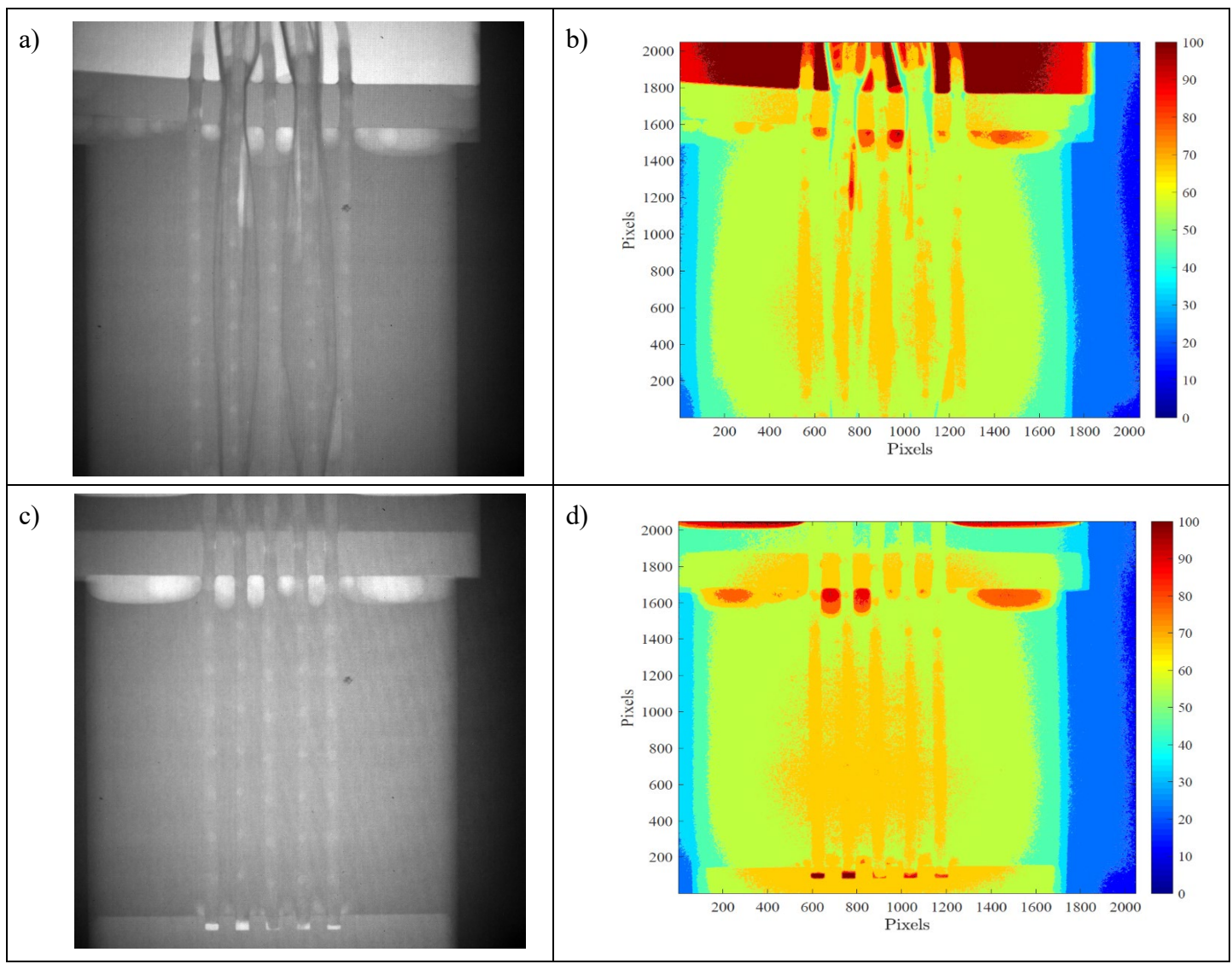

**Figure 8**

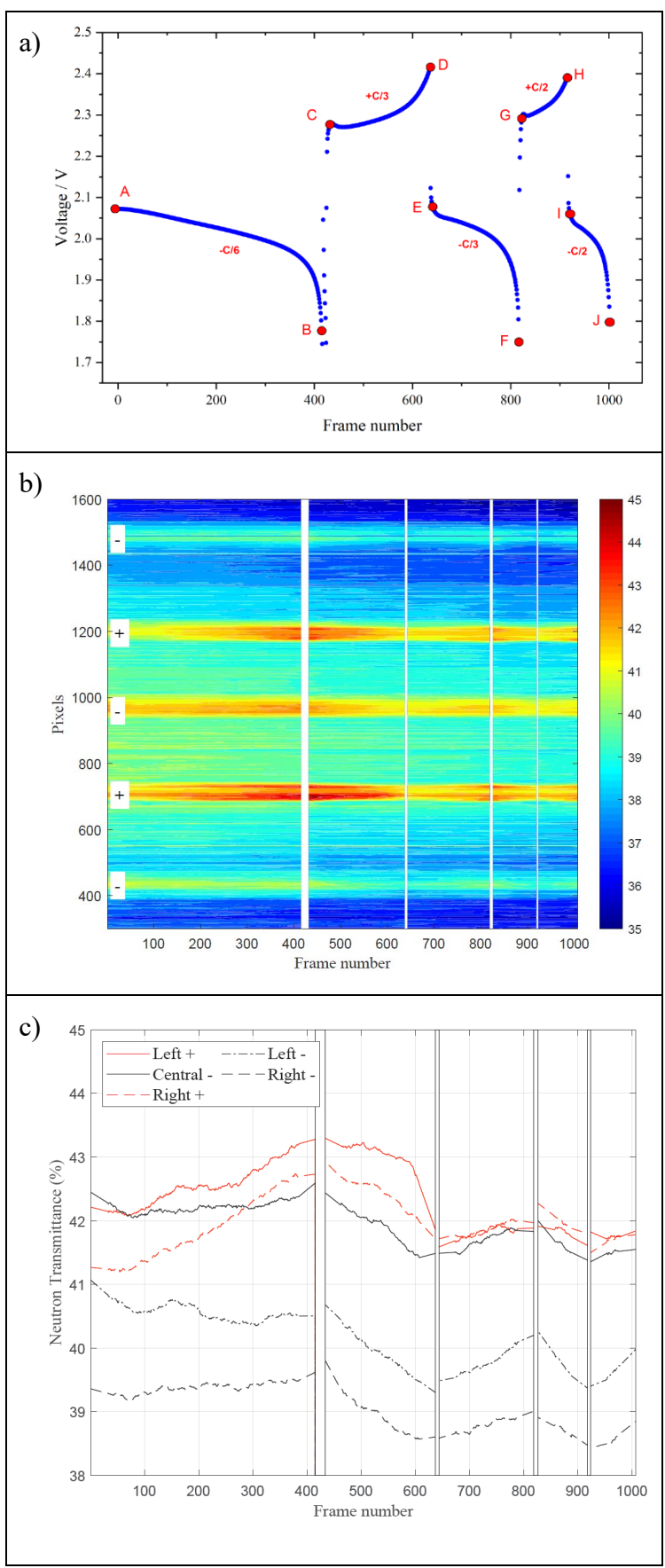

# Table captions

**Table 1.** Chemical sample information of the electrolyte components.

**Table 2.** Technical characteristics of DINGO and NECTAR facilities.

**Table 3.** Measured linear attenuation coefficients of $D_2O$ and $D_2SO_4$ and literature values.

**Table 4.** Experimental data of galvanostatic charging and discharging processes applied to the neutron friendly lead-acid battery (1.2 Ah).

**Table 1**

| Chemical name | Source | wt % in D₂O | atom % D | $\rho$ (g cm$^{-3}$) | CAS |
|---|---|---|---|---|---|
| Deuterated water | Sigma Aldrich |  | 99.9 | 1.107 | 7789-20-0 |
| Deuterated sulphuric acid | Sigma Aldrich | 96-98 | 99.5 | 1.86 | 13813-19-9 |

**Table 2**

|  | **DINGO** | **NECTAR** |
|---|---|---|
| Neutron source - Reactor | OPAL (20 MW) | FRM II (20 MW) |
| Moderator | $D_2O$ | $D_2O$ |
| Beam spectrum | thermal | fission, thermal |
| Mean energy: |  | (thermal) |
| meV | 25 | 28 |
| Å | 1.08 | 1.81 |
| Maximum neutron flux at sample (n cm$^{-2}$ s$^{-1}$) | > $10^7$ | $10^7$ |
| *L/D* | 500 or 1000 | ≤ 233 ± 16 |
| Pixel size (μm) | 27 | 13.5 |
| Detector area (cm$^2$) | 5 × 5 or 20 × 20 | 10 × 10 to 40 × 40 |
| Maximum sample size (cm$^3$) |  | 80 × 80 × 80 |

**Table 3**

|  | $\mu$ (D$_2$O) (cm$^{-1}$) | $\mu$ (D$_2$SO$_4$) (cm$^{-1}$) | $\mu$ (Pb) (cm$^{-1}$) | $\mu$ (porous Pb) (cm$^{-1}$) | $\mu$ (porous PbO$_2$) (cm$^{-1}$) | $\mu$ (PbSO$_4$) (cm$^{-1}$) |
|---|---|---|---|---|---|---|
| **DINGO** ($\lambda$ = 1.08 Å) | 0.32±0.04 | 0.22±0.03 | | | | |
| **NECTAR** ($\lambda$ = 1.81 Å) | 0.42±0.02 | 0.28±0.01 | | | | |
| **NIST** ($\lambda$ = 1.54 Å) [67] ($\rho$ (g cm$^{-3}$)) | 0.137 (1.107) | 0.051 (1.86) | 0.005 (11.34) | 0.002 (4.97) | 0.002 (5.5) | 0.008 (6.29) |
| **Kang *et. al.*** ($\lambda$ = 0.8-6 Å) [66] | 5.542 | | | | | |

**Table 4**

| Process | $I$ (A) | $Q$ (Ah) | Cut-off voltage (V) | | $V_{electrical}$ (V) | $V_{neutron}$ (V) |
|---|---|---|---|---|---|---|
| C/6 discharge | −0.156 | −4077 | 1.77 | A | 2.095 | 2.072 |
| | | | | B | 1.673 | 1.778 |
| C/3 charge | +0.272 | +3626 | 2.40 | C | 2.164 | 2.271 |
| | | | | D | 2.471 | 2.417 |
| C/3 discharge | −0.273 | −3123 | 1.70 | E | 2.111 | 2.078 |
| | | | | F | 1.598 | 1.750 |
| C/2 charge | +0.379 | +2369 | 2.40 | G | 2.164 | 2.292 |
| | | | | H | 2.468 | 2.392 |
| C/2 discharge | −0.380 | −2057 | 1.65 | I | 2.081 | 2.086 |
| | | | | J | 1.687 | 1.799 |